\documentclass[camera]{jpaper}

\usepackage[nocompress]{cite}
\usepackage{algorithmic}
\usepackage{array}

\makeatletter
\let\MYcaption\@makecaption
\makeatother

\usepackage[font=footnotesize]{subcaption}

\makeatletter
\let\@makecaption\MYcaption
\makeatother

\usepackage{fixltx2e}
\usepackage{dblfloatfix}
\usepackage[nolessnomore, italic]{mathastext}
\usepackage[T1]{fontenc}
\usepackage[usenames,dvipsnames,svgnames,table]{xcolor}
\usepackage{multirow}
\usepackage{hhline}
\usepackage[normalem]{ulem}
\usepackage{enumitem}
\usepackage{setspace}
\usepackage{indentfirst}
\usepackage{footmisc}
\usepackage{fancyhdr}
\usepackage[binary-units=true]{siunitx}
\usepackage[us,12hr]{datetime}
\usepackage[keeplastbox]{flushend}
\usepackage[hidelinks]{hyperref}

\newif\ifcameraready
\camerareadytrue

\newcommand{\versionnum}[0]{4}

\ifcameraready
\else
\fi

\ifcameraready
  \newcommand{\todo}[1][]{}
  
  \newcommand{\chII}[1]{#1}
  \newcommand{\chIII}[1]{#1}
\else
  \newcommand{\todo}[1][]{\textbf{\fcolorbox{black}{red}{\color{white}{TODO}}} \underline{$\overline{\hbox{\emph{#1}}}$}}

  \newcommand{\chII}[1]{{\color{MidnightBlue} #1}}
  \newcommand{\chIII}[1]{{\color{BrickRed} #1}}
\fi

\usepackage[normalem]{ulem}

\usepackage{epsfig}
\usepackage{graphicx}
\usepackage{dblfloatfix}
\usepackage{amsmath}
\usepackage{amssymb}
\usepackage{balance}
\usepackage{multirow}
\usepackage{url}
\usepackage{color}
\usepackage{varwidth}
\usepackage{authblk}

\begin{document}

\title{
    Read Disturb Errors in MLC NAND Flash Memory%
}

\author{Yu Cai$^1$%
\qquad%
Yixin Luo$^1$%
\qquad%
Saugata Ghose$^1$%
\qquad%
Erich F. Haratsch$^2$%
\qquad%
Ken Mai$^1$%
\qquad%
Onur Mutlu$^{3,1}$}
\affil{\it{$^1$Carnegie Mellon University%
\qquad%
$^{2}$Seagate Technology%
\qquad%
$^{3}$ETH Z\"urich}%
}

\date{}
\maketitle



\begin{abstract}

This paper summarizes our work on experimentally characterizing, mitigating,
and recovering read disturb errors in multi-level cell (MLC) NAND flash
memory, which was published in DSN 2015~\cite{cai.dsn15}, and examines the
work's significance and future potential.
NAND flash memory reliability continues to degrade as the memory is scaled down
and more bits are programmed per cell. A key contributor to this reduced
reliability is \emph{read disturb}, where a read to one row of cells impacts the
threshold voltages of \emph{unread} flash cells in different rows of the same block.
Such disturbances may shift the threshold voltages of these unread cells to
different logical states than originally programmed, leading to read errors
that hurt endurance.

For the first time in open literature, this work experimentally characterizes
read disturb errors on state-of-the-art 2Y-nm
(i.e., 20-24 nm) MLC NAND flash memory chips.  Our findings (1)
correlate the magnitude of threshold voltage shifts with read operation counts,
(2) demonstrate how program/erase cycle count and retention age affect the
read-disturb-induced error rate, and (3) identify that lowering pass-through
voltage levels reduces the impact of read disturb and extend flash lifetime.  
Particularly, we find that the  probability of read disturb errors increases with 
both higher wear-out and higher pass-through voltage levels.

We leverage these findings to develop two new techniques. The first
technique mitigates read disturb errors by dynamically tuning the pass-through
voltage on a per-block basis.  Using real workload traces, our evaluations show that this technique
increases flash memory endurance by an average of 21\%. The second technique
recovers from previously-uncorrectable flash errors by identifying and
probabilistically correcting cells susceptible to read disturb errors. Our evaluations show 
that this recovery technique reduces the raw bit error rate by 36\%.

\end{abstract}


\section{Introduction}

NAND flash memory currently sees widespread usage as a storage device, having been
incorporated into systems ranging from mobile devices and client computers to
data center storage, as a result of its increasing capacity and decreasing cost
per bit. The increasing capacity and lower cost are mainly driven by
aggressive transistor scaling and \emph{multi-level
cell} (MLC) technology, where a single flash cell can store more than one bit of
data.  However, as NAND flash memory capacity increases, flash memory suffers
from
different types of circuit-level noise, which greatly impact its
reliability.  These include program/erase cycling
noise~\cite{cai.date12, cai.date13}, cell-to-cell program interference
noise~\cite{cai.sigmetrics14, cai.date12, cai.iccd13}, retention
noise~\cite{cai.iccd12, cai.hpca15, mielke.irps08, cai.date12, cai.itj13, liu.fast12}, and
read disturb noise~\cite{cooke.fms07, mielke.irps08, takeuchi.jssc99,
grupp.micro09}.  Among all of these types of noise, \emph{read disturb} noise has largely
been understudied in the past for MLC NAND flash, with no
open-literature work available prior to our DSN 2015 paper~\cite{cai.dsn15} that characterizes and
analyzes the read disturb phenomenon.

One reason for this prior neglect has been the heretofore low occurrence of
read-disturb-induced errors in older flash technologies.  In \emph{single-level cell}
(SLC) NAND flash, read disturb errors were only expected to appear after an
average of
one million reads to a single flash block~\cite{yaffs.mitigation, grupp.micro09}.
Even with the introduction of MLC NAND flash, first-generation MLC devices were
expected to exhibit read disturb errors after 100,000
reads~\cite{yaffs.mitigation, ha.apsys13}.  As a result of manufacturing
process technology scaling, some
modern MLC NAND flash devices are now prone to read disturb errors after as few
as
20,000 reads, with this number expected to drop even further with continued
scaling~\cite{ha.apsys13, yaffs.mitigation}.  The exposure of these read disturb
errors can be exacerbated by the uneven distribution of reads across flash
blocks in contemporary workloads~\cite{umass.storagetraces,narayanan.tos08}, where certain flash blocks experience
high
temporal locality and can, therefore, more rapidly exceed the read count at
which read disturb errors are induced. We refer the reader to our prior works
for a more detailed background~\cite{cai.dsn15,cai.procieee17, cai.procieee.arxiv17, cai.arxiv17}.

\begin{figure}[h]
  \centering
  \includegraphics[trim=0 400 300 0, clip, width=.9\columnwidth]{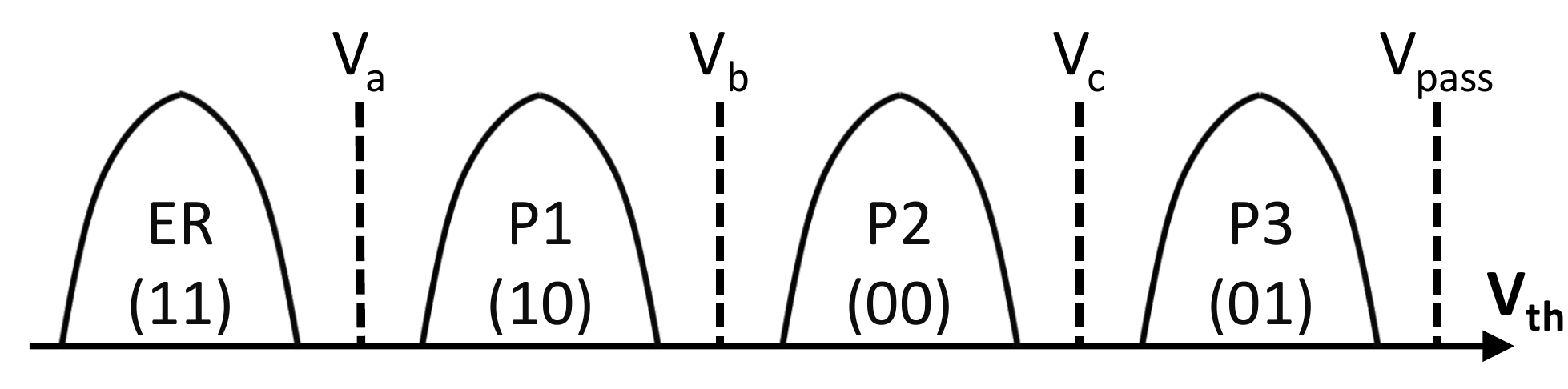}
  \caption{Threshold voltage distribution in 2-bit MLC NAND flash.  Stored data
  values are represented as the tuple (LSB, MSB). Reproduced from~\cite{cai.dsn15}.}
  \label{fig:vth}
\end{figure}

Read disturb errors are an intrinsic result of the flash architecture.  Inside
each flash cell, data is stored as the \emph{threshold voltage} of the cell,
based on the logical value that the cell represents.  As shown in
Figure~\ref{fig:vth}, during a read operation to the cell, a \emph{read
reference voltage} (i.e., $V_a$, $V_b$, or $V_c$) is applied to the transistor
corresponding to this cell.  If this read reference voltage is higher than the
threshold voltage of the cell, the transistor is turned \emph{on}.  The region
in which the threshold voltage of a flash cell falls represents the cell's
current state, which can be ER (or erased), P1, P2, or P3.  Each state decodes
into a 2-bit value that is stored in the flash cell (e.g., 11, 10, 00, or 01).
Note that the threshold voltage of all flash cells in a chip is bounded by an
upper limit, $V_{pass}$, which is the \emph{pass-through voltage}. More
detailed explanations of how NAND flash memory cells work and the data
retention errors in NAND flash memory can be found in our prior
works~\cite{cai.hpca15, cai.procieee17, cai.procieee.arxiv17, cai.arxiv17}.

Within a flash
block, the transistors of multiple cells, each from a different flash page, are
tied together as a single \emph{bitline}, which is connected to a single output
wire.  Only one cell is read at a time per bitline.  In order to read one cell
(i.e., to determine whether it is turned \emph{on} or \emph{off}), the transistors for the
cells \emph{not being read} must be kept \emph{on} to allow the value from the
cell being read to propagate to the output.  This requires the transistors to be
powered with a \emph{pass-through voltage}, which is a read reference voltage
guaranteed to be higher than \emph{any} stored threshold voltage (see
Figure~\ref{fig:vth}). Though these other
cells are \emph{not} being read, this high pass-through voltage induces
electric tunneling that can shift the threshold voltages of these unread cells to
higher values, thereby \emph{disturbing the cell contents on a read operation to
a neighboring page}.  As we
scale down the size of flash cells, the transistor oxide becomes thinner, which
in turn increases this tunneling effect.  With each read operation having an increased
tunneling effect, it takes fewer read operations to neighboring pages for the unread flash
cells to become disturbed (i.e., shifted to higher threshold voltages) and move
into a different logical state.

\begin{figure*}[t]
  \centering
  \includegraphics[trim=0 330 180 0, clip, width=\linewidth]{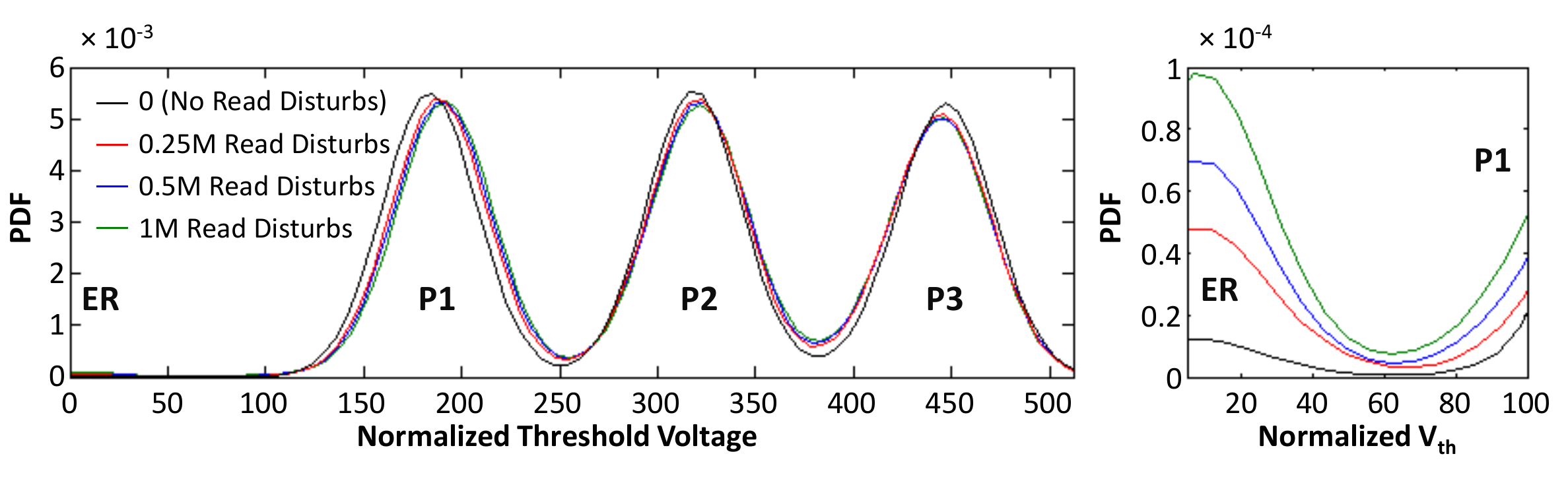}
  \caption{(a)~Threshold voltage distribution of all \chII{programmed} states before and after
    read disturb; (b)~Threshold voltage distribution between erased state and
    P1 state. Reproduced from~\cite{cai.dsn15}.
    }
  \label{fig:disturbdist}
  \vspace{10pt}
\end{figure*}


In light of the increasing sensitivity of flash memory to read disturb
errors, our goal is to (1) develop a thorough understanding of
read disturb errors in state-of-the-art MLC NAND flash memories, by performing
experimental characterization of such errors on existing commercial 2Y-nm (i.e.,
20-24 nm) flash memory chips, and (2) develop mechanisms that can tolerate
read disturb errors, making use of insights gained from our read disturb error characterization.
The \emph{key findings} from our quantitative characterization are:

\begin{itemize}

  \item The effect of read disturb on threshold voltage distributions and raw
    bit error rates increases with both the number of reads to neighboring pages and
    the number of program/erase cycles on a block.

  \item Cells with lower threshold voltages are more susceptible to errors as a
    result of read disturb.
  
  \item As the pass-through voltage decreases, (1) the read disturb effect of
    each individual read operation becomes smaller, but (2) the read errors can
    increase due to reduced ability in allowing the read value to pass through
    the unread cells.

  \item If a page is recently written, a
      significant margin within the \emph{ECC correction capability} (i.e., the
total number of bit errors it can correct for a single read) is
      unused (i.e., the page can still tolerate more
    errors), which enables the page's pass-through voltage
    to be lowered safely).

\end{itemize}

We exploit these studies on the relation between the read disturb effect and the
pass-through voltage ($V_{pass}$), to design two mechanisms that reduce the \chII{reliability}
impact of read disturb.  First, we propose a low-cost dynamic mechanism called
\emph{$V_{pass}$ Tuning}, which, for each block, finds the lowest pass-through
voltage that retains data correctness. $V_{pass}$ Tuning extends flash endurance
by exploiting the finding that a lower $V_{pass}$ reduces the read disturb error
count. Our evaluations using real workload traces show that $V_{pass}$ Tuning
extends flash lifetime by 21\%. Second, we propose \emph{Read Disturb
Recovery} (RDR), a mechanism that exploits the differences in the susceptibility of
different cells to read disturb to extend the
effective correction capability of error-correcting codes (ECC).  RDR
probabilistically identifies and corrects cells susceptible to read disturb
errors.  Our evaluations show that RDR reduces the raw bit error rate by 36\%.

\section{Characterizing Read Disturb in \\ \chII{Real NAND Flash Memory} Chips}


We use an FPGA-based NAND flash testing platform in order to characterize read
disturb on
state-of-the-art flash \chII{chips~\cite{cai.fccm11, cai.procieee17, cai.procieee.arxiv17, cai.arxiv17}}.  We use the \emph{read-retry}
operation present within MLC NAND flash devices to accurately read the cell
threshold \chII{voltage~\cite{cai.hpca15, cai.iccd12, cai.date13, park.isscc11,
cai.iccd13, cai.sigmetrics14, cai.hpca17, fukami2017improving,
cai.procieee17, cai.procieee.arxiv17, cai.arxiv17}}. As
threshold voltage values are proprietary information, we present our results
using a \emph{normalized threshold voltage}, where the nominal value of
$V_{pass}$ is equal to 512 in our normalized scale, and where 0 represents
GND.

One limitation of using commercial flash devices is the inability to alter
the $V_{pass}$ value, as no such interface currently exists.  We work around
this by using the read-retry mechanism, which allows us to change the read
reference voltage $V_{ref}$ one wordline at a time.  Since both $V_{pass}$ and $V_{ref}$ are applied
to wordlines, we can mimic the effects of changing $V_{pass}$ by instead
changing $V_{ref}$ and examining the impact on the wordline being read.
We perform these experiments on one 
wordline per block, and repeat them over ten different blocks.

We present our major findings below. For a complete description of all of our
observations, we refer the reader to our DSN 2015 paper~\cite{cai.dsn15}.

\subsection{Quantifying Read Disturb Perturbations}


First, we quantify the amount by which read disturb shifts the threshold
voltage, by measuring threshold voltage values for unread cells after 0, 250K, 
500K, and 1~million read operations to other cells within the same flash block.
Figure~\ref{fig:disturbdist}a shows the distribution of the threshold voltages for
cells in a flash block after 0, 250K, 500K, and 1 million read operations.
Figure~\ref{fig:disturbdist}b zooms in on this to illustrate the distribution 
for values in the ER
state.
We find that \emph{the magnitude of the threshold voltage shift for a
cell due to read disturb (1) increases with the number of read disturb
operations, and (2) is higher if the cell has a lower threshold voltage.}

\subsection{Effect of Read Disturb on Raw Bit Error Rate}

\begin{figure}[t]
  \centering
  \includegraphics[trim=12 370 445 0, clip, width=\linewidth]{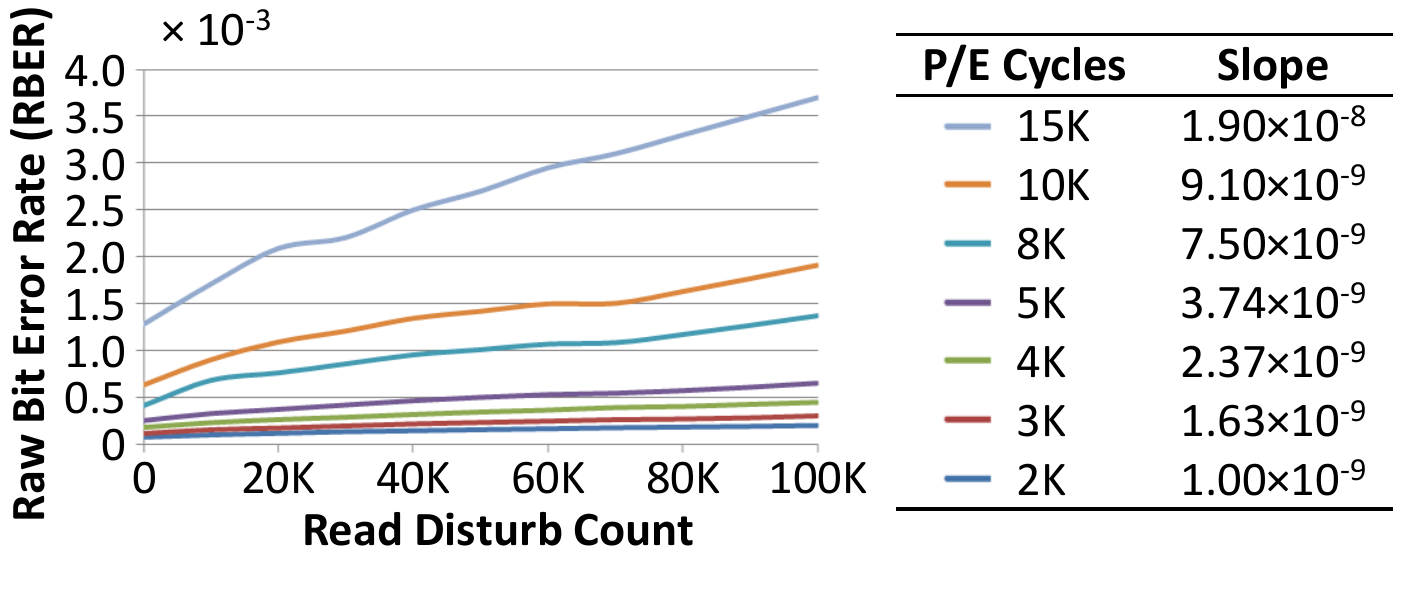}%
  \caption{Raw bit error rate vs. read disturb count under
  different levels of program and erase (P/E) wear. Reproduced
  from~\cite{cai.dsn15}.}
  \label{fig:disturbrber}
\end{figure}

Second, we aim to relate these threshold voltage shifts to the \emph{raw bit error rate} (RBER),
which refers to the probability of reading an incorrect state from a flash cell. We measure whether flash cells that are more worn out (i.e., cells that
have been programmed and erased more times) are impacted differently due to
read disturb.  Figure~\ref{fig:disturbrber} shows the RBER over an increasing
number of read disturb operations for different amounts of P/E cycle wear 
(i.e., the amount of wearout in P/E cycles) on
flash blocks.  Each level shows a linear RBER increase as the read
disturb count increases.  We find that \emph{(1)~for a given amount of P/E cycle wear on a block, the raw bit
error rate increases roughly linearly with the number of
read disturb operations}, and that \emph{(2)~the effects of read disturb are greater for cells
that have experienced a larger number of P/E cycles.}

\subsection{Pass-Through Voltage Impact on \\ Read Disturb}

\begin{figure}[t]
  \centering
  \includegraphics[trim=5 285 260 0, clip, width=\linewidth]{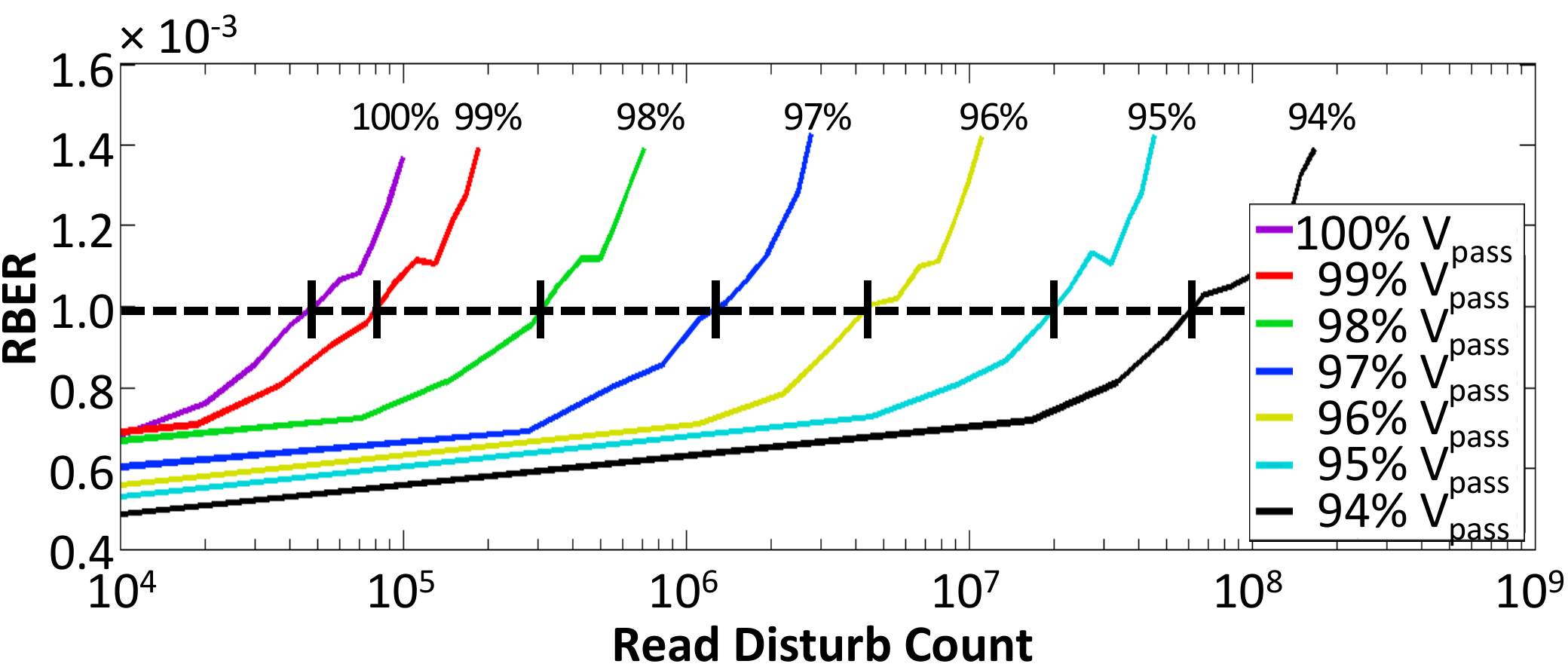}
  \caption{Raw bit error rate vs. read disturb count for different
  $V_{pass}$ values, for flash memory under 8K program/erase cycles of wear.
  Reproduced from~\cite{cai.dsn15}.}
  \label{fig:readdisturbvpassrber}
\end{figure}

Third, we show that the cause of read disturb can be
reduced by \chII{reducing (i.e., relaxing)} the pass-through voltage using
a \chII{circuit-level model of the flash cell}, and verify
this observation using real measurements. Figure~\ref{fig:readdisturbvpassrber}
shows the measured change in RBER as a function of
the number of read operations, for selected relaxations of $V_{pass}$.  Note
that the x-axis uses a log scale.  \emph{For a fixed number of reads, even a
small decrease in the $V_{pass}$ value can yield a significant decrease in
RBER.}  As an example, at 100K reads, lowering $V_{pass}$ by 2\% can reduce the
RBER by as much as 50\%.  Conversely, \emph{for a fixed RBER, a decrease in $V_{pass}$
exponentially increases the number of tolerable read disturbs.} However, decreasing $V_{pass}$ can prevent some
\chII{cells'} values from propagating correctly along the bitline on a read, as an
unread flash cell transistor may be incorrectly turned off, thus generating new errors.  Unlike read disturb errors,
these bitline propagation errors (or \emph{read errors}) \emph{do not alter the
threshold voltage
of the flash cell}.

\subsection{Effect of Pass-Through Voltage on \\ Raw Bit Error Rate}

Fourth, setting $V_{pass}$ to a value slightly lower than the maximum $V_{th}$
leads to a trade-off. On the one hand, it can substantially reduce the effects
of read disturb.  On the other hand, it causes a small number of unread cells
to incorrectly stay \emph{off} instead of passing through a value, potentially
leading to a \emph{read error}. Therefore, if the number of read disturb errors
can be dropped significantly by lowering $V_{pass}$, the small number of read
errors introduced may be warranted. If too many read errors occur, we can
always fall back to using the maximum threshold voltage for $V_{pass}$ without
consequence. Naturally, this trade-off
depends on the magnitude of these error rate changes. We now explore the gains
and costs, in terms of overall RBER, for relaxing $V_{pass}$ \emph{below} the
maximum threshold voltage of a block.

\begin{figure}[h]
  \centering
  \includegraphics[trim=0 315 398 0, clip, width=0.95\linewidth]
  {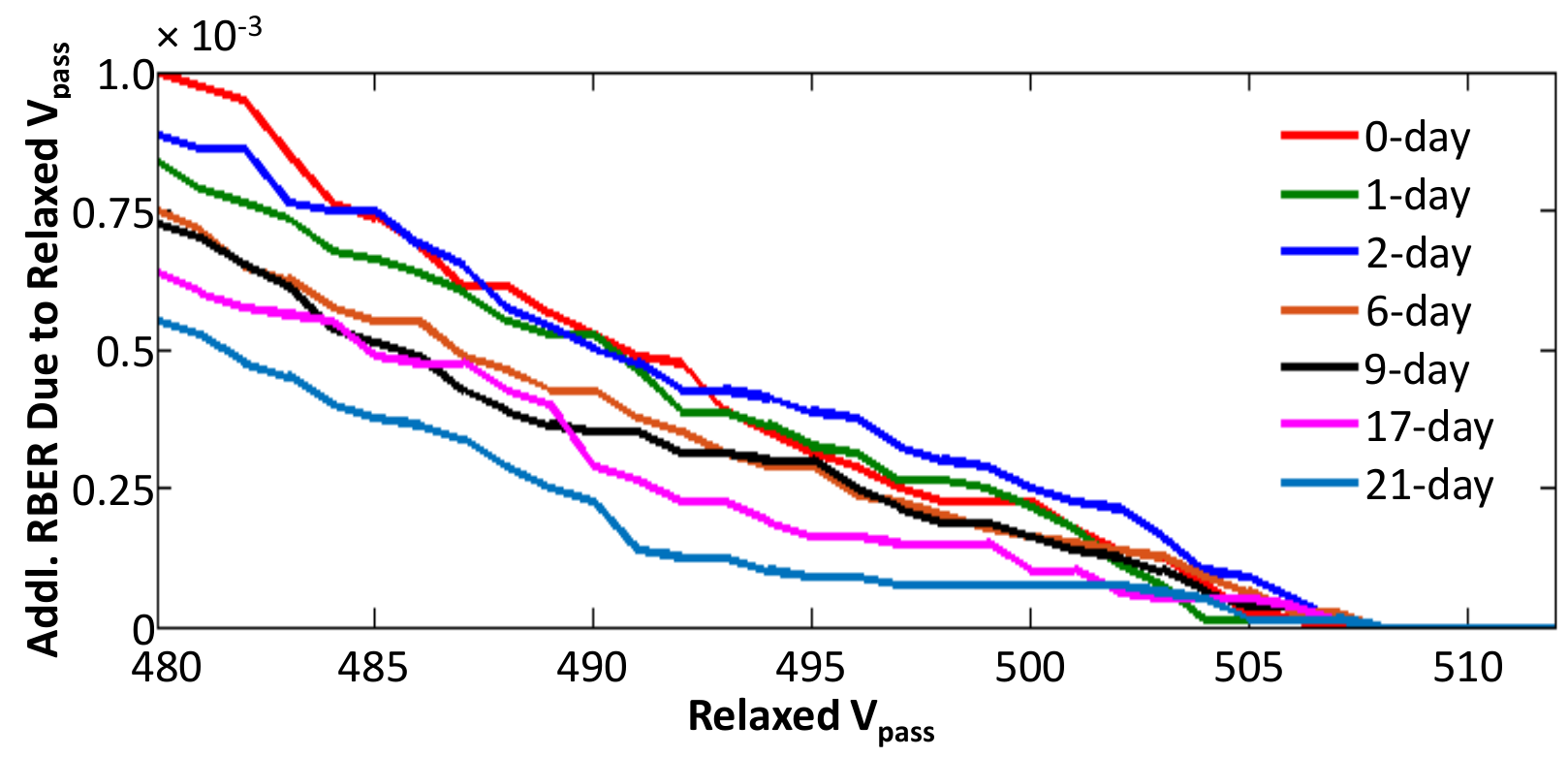}
  \caption{Additional raw bit error rate induced by relaxing $V_{pass}$, shown
  across a range of data retention ages. Reproduced
  from~\cite{cai.dsn15}.}
  \label{fig:relaxedvpassrber}
\end{figure}

To identify the extent to which relaxing $V_{pass}$ affects the raw bit error
rate, we experimentally sweep over $V_{pass}$, reading the data after a range of
different retention ages, as shown in Figure~\ref{fig:relaxedvpassrber}. 
First,
we observe that across all of our studied retention ages, \emph{$V_{pass}$ can
be lowered to some degree without inducing any read errors}.  For greater
relaxations, though, the error rate increases as more unread cells are
incorrectly turned \emph{off} during read operations.  We also note that, \emph{for a
given $V_{pass}$ value, the additional read error rate is lower if the read is
performed a longer time after the data is programmed into the flash (i.e., if
the retention age is longer).}  This is because of the retention loss effect,
where cells slowly leak charge and thus have lower threshold voltage values over
time.  Naturally, as the threshold voltage of every cell decreases, a relaxed
$V_{pass}$ becomes more likely to correctly turn \emph{on} the unread cells.

\subsection{Error Correction with Reduced \\ Pass-Through
Voltage}

\begin{figure}[h]
  \includegraphics[trim=0 295 380 20, clip, width=\linewidth]{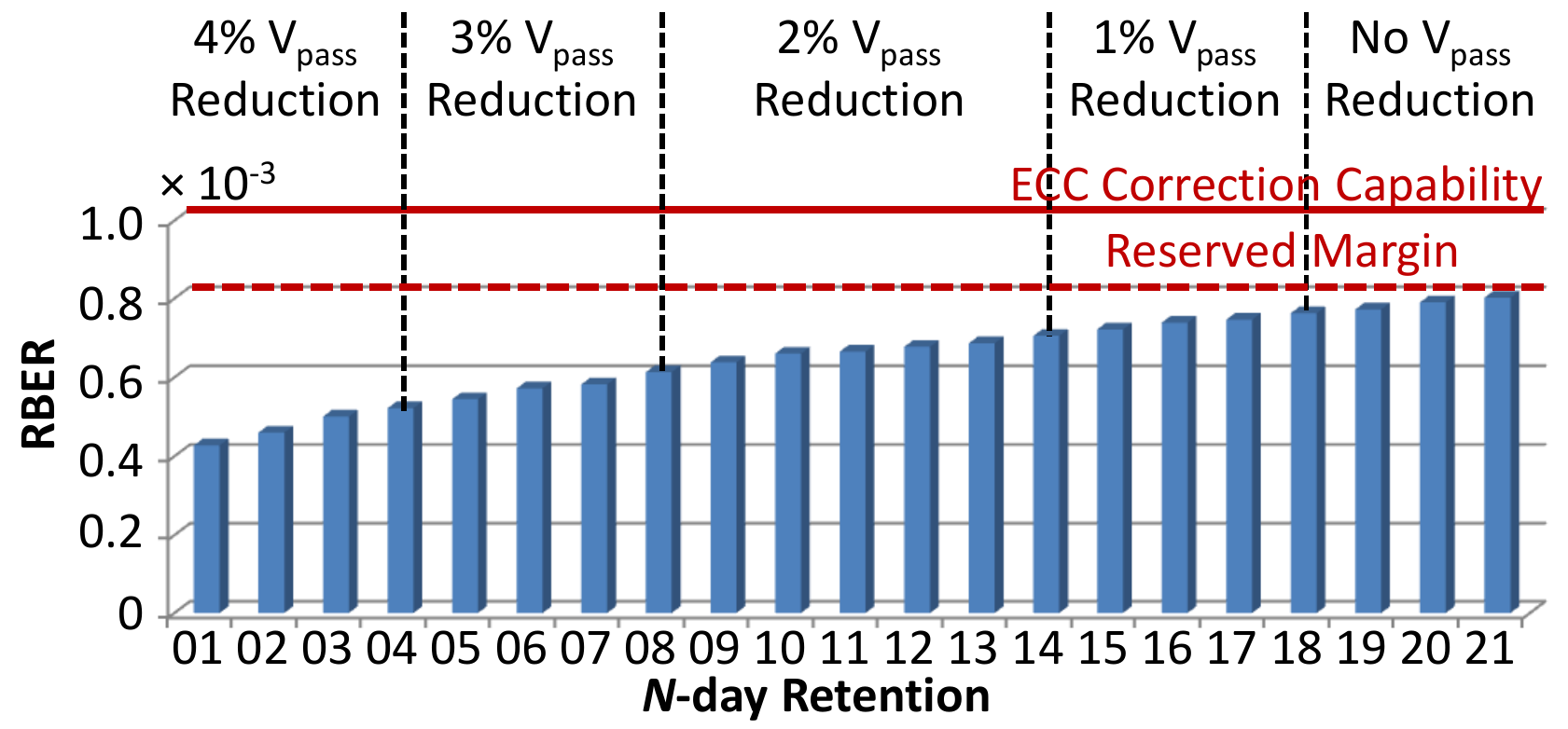}
  \caption{Overall raw bit error rate and tolerable $V_{pass}$
    reduction vs.
  retention age, for a flash block with 8K P/E cycles of wear. Reproduced
  from~\cite{cai.dsn15}.}
  \label{fig:ecccapability}
\end{figure}

Fifth, while we have shown, in Section~3.6 of our DSN 2015 paper~\cite{cai.dsn15},
that $V_{pass}$ can be lowered to some
degree without introducing new raw bit errors, we would ideally like to further
decrease $V_{pass}$ to lower the read disturb impact more.  This can enable
flash devices to tolerate many more reads. The ECC used for NAND flash
memory can tolerate an RBER of up to 10\textsuperscript{--3}~\cite{cai.iccd12,
cai.itj13}, which occurs only during worst-case conditions such as long
retention time. Our goal is to identify how many additional raw bit errors
the current level of ECC provisioning in flash chips can sustain.
Figure~\ref{fig:ecccapability} shows how the expected RBER changes over a
21-day period for our tested flash chip \emph{without read disturb}, using a
block with 8,000 P/E cycles of wear. An RBER margin
(20\% of the total ECC correction capability) is reserved to
account for variations in the distribution of errors and other potential errors
(e.g., program and erase errors). For each retention age,
the maximum percentage of \emph{safe} $V_{pass}$ reduction (i.e., the lowest
value of $V_{pass}$ at which all read errors can still be corrected by ECC)
is listed on the top of Figure~\ref{fig:ecccapability}.
As we can see, by exploiting the previously-unused ECC correction capability,
$V_{pass}$ can be safely reduced by as much as 4\% when the retention age is low
(less than 4 days).

Our key insight from this study is that \emph{a lowered $V_{pass}$ can
reduce the effects of read disturb, and that the read errors induced from
lowering $V_{pass}$ can be tolerated by the built-in error correction mechanism
within modern flash controllers}. More results and more detailed analysis are in
our DSN 2015 paper~\cite{cai.dsn15}.

\section{Mitigation: Pass-Through Voltage Tuning}

\begin{figure}[h]
  \centering
  \includegraphics[width=.9\linewidth]{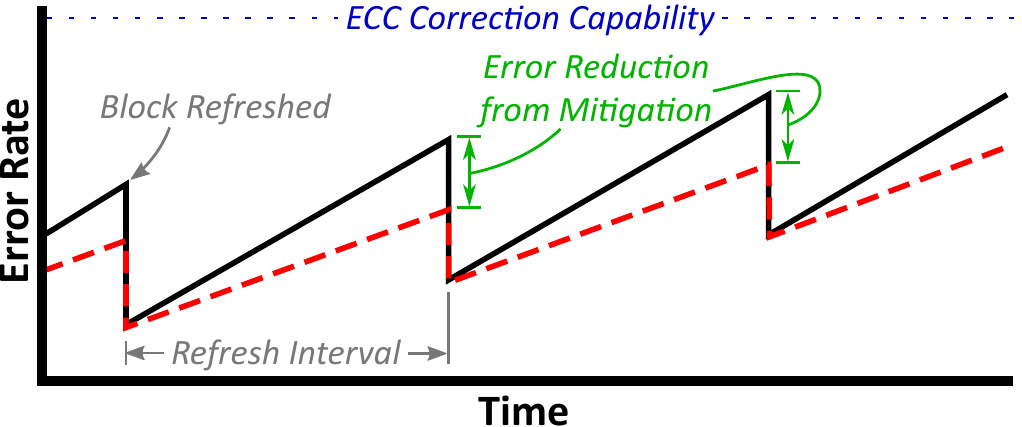}
  \caption{Exaggerated example of how read disturb mitigation reduces
    error rate peaks for each refresh interval.  Solid black line is the unmitigated
    error rate, and dashed red line is the error rate after mitigation.  (Note that
    the error rate does not include read errors introduced by reducing $V_{pass}$, as
    the unused error correction capability can tolerate errors caused by
  \emph{$V_{pass}$ Tuning}.) Reproduced
  from~\cite{cai.dsn15}.}
  \label{fig:errorcycle}
\end{figure}

To minimize the effect of read disturb, we propose a mechanism called \emph{$V_
{pass}$ Tuning}, which \emph{learns the minimum
pass-through voltage for each block, such that all data
within the block can be read correctly with ECC}. 
Figure~\ref{fig:errorcycle} provides an exaggerated illustration of how the
unused ECC capability changes over the retention period (i.e., the \emph{refresh
interval}).  At the start of each retention period, there are no retention
errors or read disturb errors, as the data has just been restored.  In
these cases, the large unused ECC capability allows us to design an
\emph{aggressive} read disturb mitigation mechanism, as we can safely
\emph{introduce correctable errors}.  Thanks to read disturb mitigation, we can
reduce the effect of each individual read disturb, thus lowering the total
number of read disturb errors accumulated by the end of the refresh interval.
This reduction in read disturb error count leads to lower \emph{error
count peaks} at the end of each refresh interval, as shown in
Figure~\ref{fig:errorcycle} by the distance between the solid black line and the
dashed red line. Since flash lifetime is dictated by the number of data errors
(i.e., when the total number of errors exceeds the ECC correction capability,
the flash device has reached the end of its life), lowering the error count
peaks extends lifetime by extending the time before these peaks exhaust the
ECC correction capability.

Our learning mechanism works
online and is triggered on a daily basis.  \emph{$V_{pass}$ Tuning} can be fully
implemented within the \emph{flash controller}, and has two components:
\begin{enumerate}[label=\arabic*.]

  \item It first finds the size of the ECC margin $M$ (i.e., the unused
    correction capability within ECC) that can be exploited to tolerate additional
    read errors for each block.  In order to do this, our mechanism discovers
    the page with \emph{approximately} the highest number of raw bit errors.

  \item Once it knows the available margin $M$, our mechanism calibrates the
    pass-through voltage $V_{pass}$ \emph{on a per-block basis} to find the
    lowest value of $V_{pass}$ that introduces no more than $M$ additional raw
    errors.

\end{enumerate}

The first component of our mechanism must first approximately discover the page
with the highest error count, which we call the \emph{predicted worst-case
page}. After manufacturing, we statically find the predicted worst-case page by
programming pseudo-randomly generated data to each page within the block, and
then immediately reading the page to find the error count, as prior work on
error analysis has done~\cite{cai.date12}. For each block, we record the page
number of the page with the highest error count. Our mechanism obtains the error
count, which we define as our \emph{maximum estimated error} ($MEE$), by
performing \emph{a single read} to this page and reading the error count
provided by ECC (once a day). We conservatively reserve 20\% of the spare ECC
correction capability in our calculations.  Thus, if the maximum
number of raw bit errors correctable by ECC is $C$, we calculate the
available ECC margin for a block as $M = (1 - 0.2) \times C - MEE$.

The second component of our mechanism identifies the greatest $V_{pass}$
reduction
that introduces no more than $M$ raw bit errors.  The general
\emph{$V_{pass}$ identification process} requires three steps:

\vspace{3pt}
\noindent\emph{Step~1}: Aggressively reduce $V_{pass}$ to $V_{pass} - \Delta$, where
$\Delta$ is the smallest resolution by which $V_{pass}$ can change.

\vspace{3pt}
\noindent\emph{Step~2}:
Apply the new $V_{pass}$ to \emph{all} wordlines in the block. Count the number
of 0's read from the page (i.e., the number of bitlines incorrectly switched
\emph{off}) as $N$.
If $N \leq M$ (recall that $M$ is the extra available ECC correction margin), 
the read errors resulting from this $V_{pass}$ value can be corrected by
ECC, so we repeat Steps~1 and~2 to try to further reduce $V_{pass}$. If $N > M$, it means we
have reduced $V_{pass}$ too aggressively, so we proceed to Step~3 to roll back
to an acceptable value of $V_{pass}$.

\vspace{3pt}
\noindent\emph{Step~3}: Increase $V_{pass}$ to $V_{pass} + \Delta$, and verify that the
introduced read errors can be corrected by ECC (i.e., $N \leq M$). If this verification
fails, we repeat Step~3 until the read errors are reduced to an acceptable
range.

\vspace{3pt}
The implementation can be simplified greatly in practice, as the error rate
changes are relatively slow over time. Over the course of the seven-day refresh
interval, our mechanism must perform one of two actions each day:

\vspace{3pt}
\noindent\emph{Action~1}:
When a block is \emph{not} refreshed, our mechanism checks once daily if $V_{pass}$ should
\emph{increase}, to accommodate the slowly-increasing number of errors due to
dynamic factors (e.g., retention errors, read disturb errors).  \vspace{3pt}

\noindent\emph{Action~2}:
When a block is refreshed, all retention and read disturb errors accumulated
during the previous refresh interval are corrected. At this time, our mechanism checks how
much $V_{pass}$ can be \emph{lowered} by.

Our mechanism repeats the $V_{pass}$ identification process for each block that
contains valid data to learn the \emph{minimum pass-through voltage we can
use}. It also repeats the entire $V_{pass}$ learning process
daily to adapt to threshold voltage changes due to retention
loss~\cite{cai.sigmetrics14, cai.iccd13}. As such, the pass-through voltage of
\emph{all blocks} in a flash drive can be fine-tuned \emph{continuously} to
reduce read disturb and thus improve overall flash lifetime. Our DSN 2015
paper~\cite{cai.dsn15}
describes this mechanism in more detail, and discusses a fallback
mechanism
for extreme cases where the additional errors
accumulating between tunings
exceed our 20\% margin of unused error correction capability.
\chII{For more detail, we refer the reader to Section~4 of our DSN 2015 paper~\cite{cai.dsn15}.}


Our mechanism can reduce $V_{pass}$ by as much as 4\%.
Through a series of optimizations, described in more detail in Section~4 of our
DSN 2015 paper~\cite{cai.dsn15}, 
it only incurs an average daily performance overhead of 24.34~sec for a 512GB SSD,
and uses only 128KB storage overhead to record per-block data.

\begin{figure}[h]
  \centering
  \includegraphics[trim=0 290 145 0, clip, width=\linewidth]{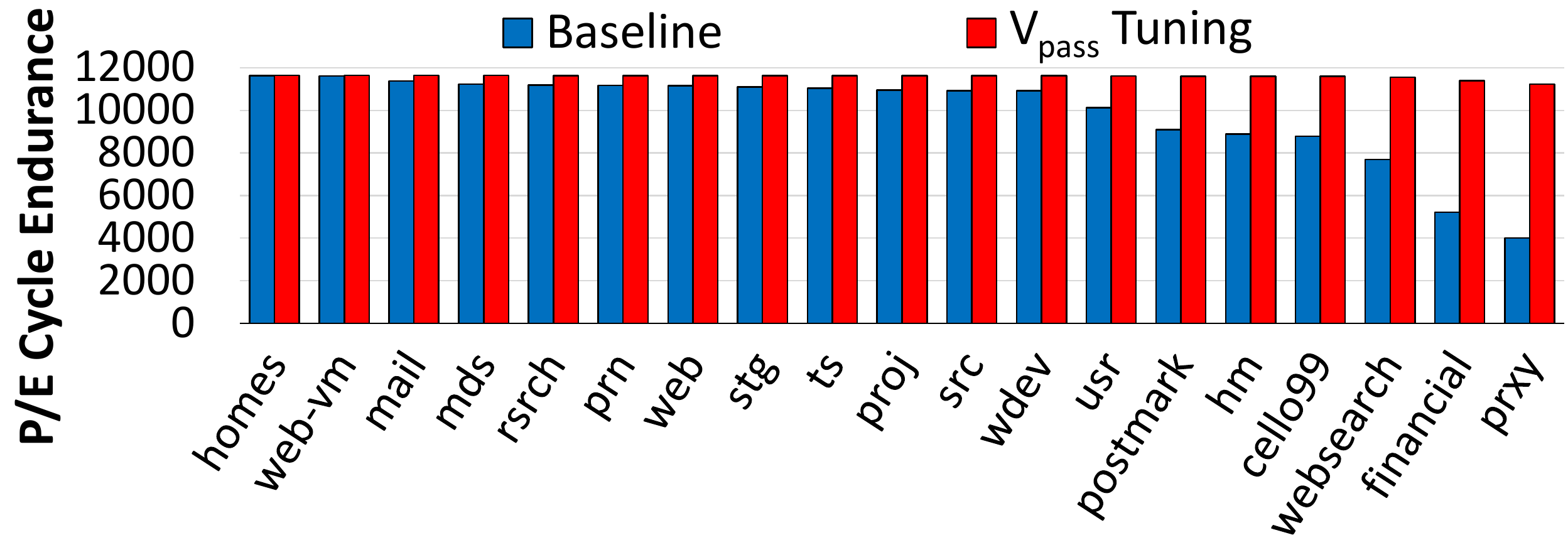}
  \caption{Endurance improvement with \emph{$V_{pass}$ Tuning}. Reproduced
  from~\cite{cai.dsn15}.}
  \label{fig:mitigationresult}
\end{figure}

We evaluate \emph{$V_{pass}$ Tuning} with I/O traces
collected from a wide range of real workloads with different use
cases~\cite{umass.storagetraces, katcher.tr97, koller.tos10, narayanan.tos08,
hplabs.teslatraces}.
Figure~\ref{fig:mitigationresult} shows how our mechanism can increase the endurance (measured as the
number of program/erase cycles that take place before the NAND flash memory can
no longer
be used).  We find that for \chII{a variety of} our workloads, our $V_{pass}$
tuning mechanism \chII{increases} flash \chII{memory endurance} by an average of 21.0\%, thanks
to its success in reducing the number of raw bit errors that occur
due to read disturb.

\section{Read Disturb Oriented Error Recovery}

Even \chII{if we mitigate} the impact of read disturb errors, a flash device will
eventually exhaust its lifetime.  At that point, some reads will have more raw
errors to correct than can be corrected by ECC, preventing the drive from
returning the \chII{correct} data to the user.  Traditionally, this \chII{is
referred to as} the point of
\emph{data loss}.

We propose to take advantage of our understanding of read disturb behavior, by
designing a mechanism that can recover data \emph{even after the device has
exceeded its lifetime}.  This mechanism, which we call \emph{Read Disturb
Recovery} (RDR), (1)~identifies flash cells that are \emph{susceptible} to generating
errors due to read disturb (i.e., \emph{disturb-prone} cells), and
(2)~\chII{probabilistically} corrects \chII{the data
stored in these cells} without the assistance of
ECC.  After \chII{these probabilistic} corrections, the number of errors
for a read will be brought back down, to a point at which ECC can successfully
correct \chII{the} remaining errors and return valid data to the user.

\begin{figure}[h]
  \centering
  \includegraphics[trim=0 390 320 0, clip, width=\linewidth]{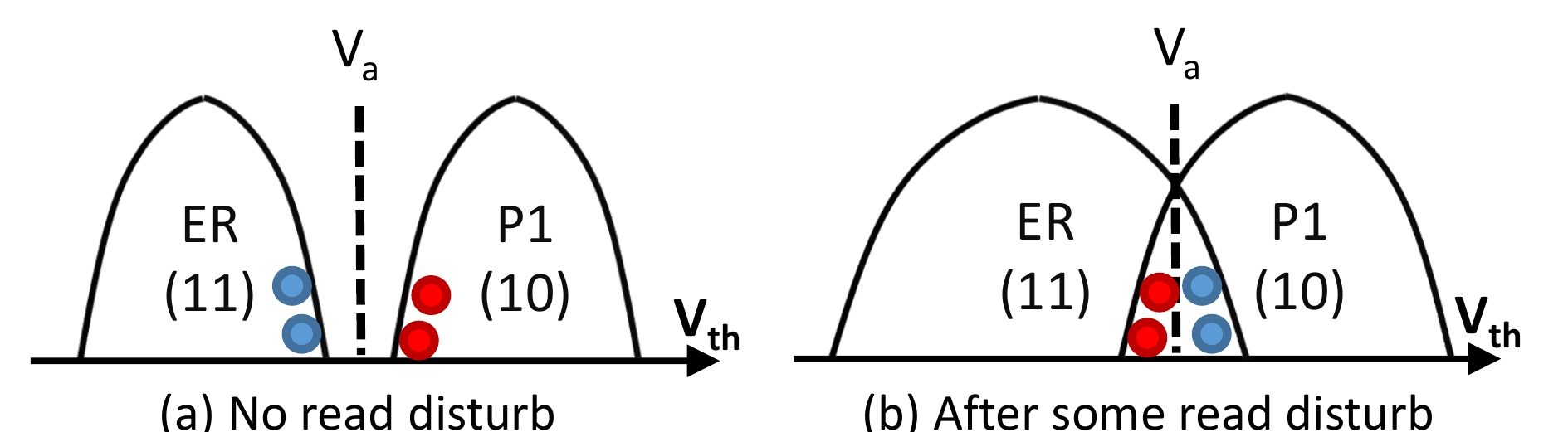}
  \caption{$V_{th}$ distributions before and after read disturb. Reproduced
  from~\cite{cai.dsn15}.}
  \label{fig:disturbfailure}
\end{figure}

To understand why \chII{identifying disturb-prone cells can help with correcting errors}, we study
why read disturb errors occur to begin with.  Figure~\ref{fig:disturbfailure}a
shows the state of four flash cells before read disturb happens.  The two blue
cells are both programmed with a two-bit value of 11, and the two red cells are
programmed with a two-bit value of 00, with each two-bit value being assigned
to a different range of threshold voltages ($V_{th}$).  Between each assigned
range is a margin.  When read disturb occurs, the blue cells, which are
\emph{disturb-prone}, experience large $V_{th}$ shifts upwards, while
the blue cells, which are \emph{disturb-resistant}, do not shift much, as
shown in Figure~\ref{fig:disturbfailure}b.  Now that the distributions of these
two-bit values overlap, a read error will occur for these four cells.

\chII{
\textbf{Identifying Susceptible Cells.} In order to identify susceptible
cells, RDR induces a significant number of additional read disturbs (e.g.,
100K) within \chII{the flash cells
that contain uncorrectable errors}. We do this by characterizing the degree of
the threshold voltage shift ($\Delta V_{th}$) induced by the additional read
disturbs, and comparing the shift to a delta threshold voltage ($\Delta
V_{ref}$) at the intersection of the two probability
density functions. We classify cells with a higher threshold voltage change ($\Delta
V_{th} > \Delta V_{ref}$) as \emph{disturb-prone} cells. We classify cells
with a lower or negative threshold voltage change ($\Delta V_{th} < \Delta
V_{ref}$) as \emph{disturb-resistant} cells. Section~5.2 of our DSN 2015
paper~\cite{cai.dsn15} provides more detailed results and analysis of
disturb-prone and disturb-resistant cells.

\textbf{Correcting Susceptible Cells.}
For flash cells with threshold voltages close to the boundary between two
different data values, RDR predicts that the disturb-prone cells belong to
the lower of the two voltage distributions (ER in
Figure~\ref{fig:disturbfailure}).  Likewise, disturb-resistant cells near the
boundary likely belong to the higher voltage distribution (P1 in 
Figure~\ref{fig:disturbfailure}).
This \chIII{does \emph{not}} eliminate all errors, but \chIII{decreases} the raw bit
errors in disturb-prone cells. RDR attempts to correct the remaining raw bit
errors using ECC. Section~5.3 of our DSN 2015 paper~\cite{cai.dsn15} provides
more detail on the RDR mechanism.}

\begin{figure}[h]
  \centering
  \includegraphics[trim=8 355 395 0, clip, width=0.88\linewidth]{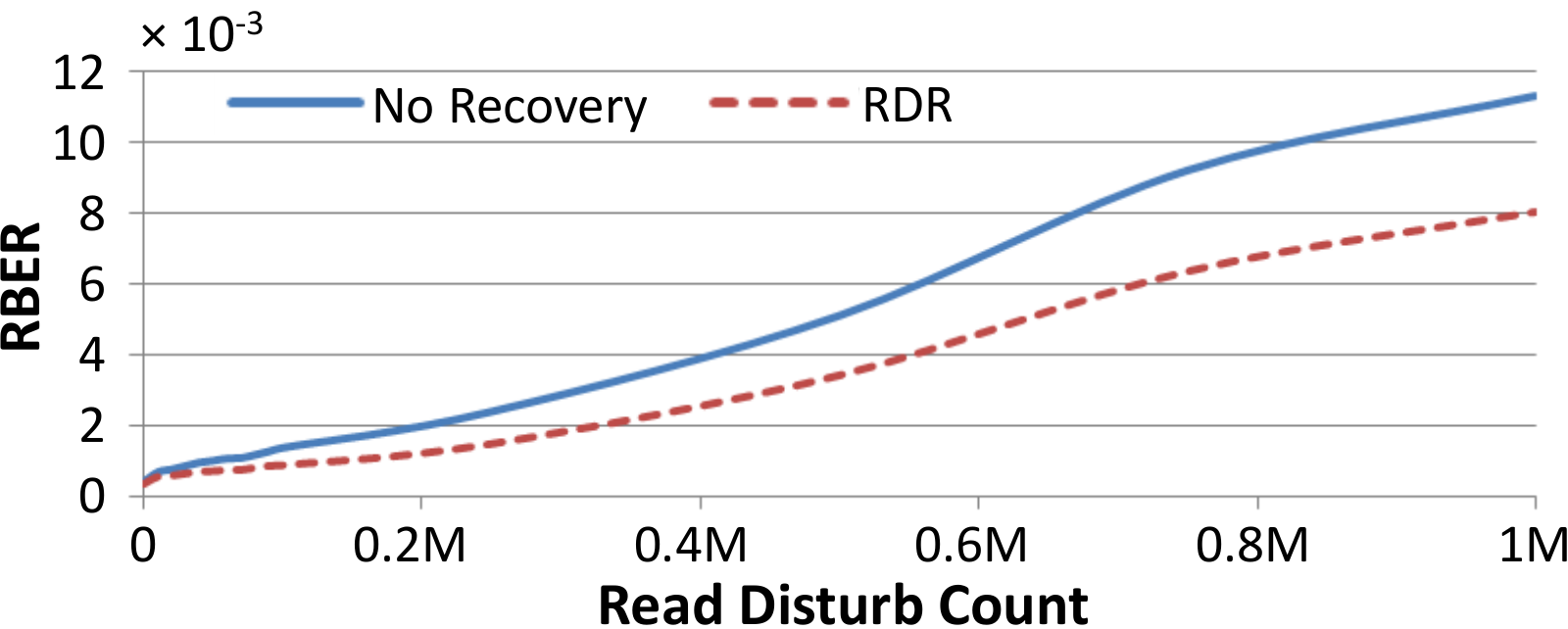}
  \caption{Raw bit error rate vs.\ number of read disturb operations, with and without
  RDR, for a flash block with 8,000 P/E cycles of wear. Reproduced
  from~\cite{cai.dsn15}.}
  \label{fig:recoveryresult}
\end{figure}

We evaluate how the overall RBER changes when we use RDR\@.
Fig.~\ref{fig:recoveryresult} shows experimental results for error recovery in
a flash block with 8,000 P/E cycles of wear. When RDR is applied, the
\emph{reduction in overall RBER grows with the read disturb count, from a few
percent for low read disturb counts} up to 36\% for 1 million read disturb
operations. As data experiences a greater number of read disturb operations,
the read disturb error count contributes to a significantly larger portion of
the total error count, which our recovery mechanism targets and reduces.  We
therefore conclude that RDR can provide a large effective extension of the ECC
correction capability.

\section{Related Work}

We \chII{break down} related work \chII{on NAND flash memory  (Section~\ref{sec:related:flash})} into six major categories: (1)~read disturb error
characterization, (2)~NAND flash memory error characterization, (3)~3D NAND
error characterization, (4)~read disturb error mitigation, (5)~voltage
optimization, and (6)~error recovery. \chII{We then introduce related work on read disturb
errors in DRAM (Section~\ref{sec:related:dramreaddisturb})
and emerging memory technologies (Section~\ref{sec:related:emerging}).}

\subsection{Related Works on NAND Flash Memory}
\label{sec:related:flash}

\textbf{Read Disturb Error Characterization.} Prior to this
work~\cite{cai.dsn15}, the read disturb phenomenon for NAND flash memory has not been
well explored in openly-available literature. \chII{Prior work~\cite{kim.isca14}
experimentally characterizes and proposes solutions for read disturb errors in
DRAM. The mechanisms for disturbance and techniques to
mitigate them are different between DRAM and NAND flash due to device-level
differences~\cite{mutlu.date17}.} Recent work has characterized concentrated read disturb
effect
and find that there are more read disturb errors on the direct neighbors to the
page being repeatedly read~\cite{zambelli2017uniform}. Recent work has found
that read disturb errors significantly reduce the reliability of unprogrammed
and partially-programmed wordlines within a flash block, and can cause security
vulnerabilities~\cite{papandreou2016effect, cai.hpca17}. These unprogrammed and
partially programmed wordlines have lower threshold voltages  (e.g., all cells
in unprogrammed wordlines are in erased state), they are more sensitive to read
disturb effect. When the wordlines are fully-programmed, NAND flash memory chip
cannot correct any of these read disturb errors and thus program the misread
flash cells into an incorrect state.

\textbf{NAND Flash Memory Error Characterization.} \chII{There are many
past works from us and other research groups that analyze many} different types
of NAND flash memory errors in MLC, planar NAND flash
memory, including P/E cycling errors~\cite{mielke.irps08, cai.date13,
parnell.globecom14, luo.jsac16, papa.glsvlsi2014}, programming
errors~\cite{cai.hpca17, parnell.globecom14, luo.jsac16}, cell-to-cell program
interference errors~\cite{cai.iccd13, cai.date13, cai.sigmetrics14}, retention
errors~\cite{mielke.irps08, cai.date13, cai.iccd12, cai.hpca15,
papa.glsvlsi2014}, and read disturb errors~\cite{mielke.irps08, cai.dsn15,
papa.glsvlsi2014}, \chII{and propose many different mitigation mechanisms.
These works complement our DSN 2015 paper. A survey of these works (and many
other related ones) can be found in our recent works~\cite{cai.procieee17, cai.procieee.arxiv17, cai.arxiv17}.} These
works characterize how raw bit error rate and
threshold voltage change over various types of noise. Our recent work
characterizes the same types of errors in TLC, planar NAND flash memory and
has similar findings~\cite{cai.procieee17, cai.procieee.arxiv17, cai.arxiv17}. Thus, we believe that most of the
findings on MLC NAND flash memory can be generalized to any types of planar
NAND flash memory devices (e.g., SLC, MLC, TLC, or QLC).
Recent work has also studied SSD errors in the field, and has shown the
system-level implications of these errors to large-scale
data centers~\cite{meza.sigmetrics15, narayanan.systor16, schroeder2017reliability}.

\textbf{3D NAND Error Characterization.} Recently, manufacturers
have begun to produce SSDs that contain \emph{three-dimensional} (3D) NAND
flash memory~\cite{yoon.fms15, park.jssc15, kang.isscc16, im.isscc15,
micheloni.procieee17, micheloni.sn16}.  In 3D NAND flash memory,
\emph{multiple layers} of flash cells are stacked vertically to increase the
density and to improve the scalability of the memory~\cite{yoon.fms15}. In
order to achieve this stacking, manufacturers have changed a number of
underlying properties of the flash memory design. However, the
internal organization of a flash block remains unchanged. Thus, read
disturb errors are similar in 3D NAND flash memory. But the rate of read
disturb errors are significantly reduced in today's 3D NAND because it
currently uses a larger manufacturing process technology~\cite{
grossi.springer16, tressler.fms15}. We refer the reader to our
prior work for a more detailed comparison between 3D NAND and planar
NAND~\cite{cai.procieee17, cai.procieee.arxiv17, cai.arxiv17}.
Recent work characterizes the latency and raw bit error
rate of 3D NAND devices based on floating gate cells~\cite{xiong.sigmetrics17}
and make similar observations as in planar NAND devices based on floating
gate cells.
Recent work has reported several differences between 3D NAND and planar NAND
through circuit level measurements. These differences include 1)~smaller
program variation at high P/E cycle~\cite{park.jssc15}, 2)~smaller program
interference~\cite{park.jssc15}, 3)~early retention loss~\cite{mizoguchi.imw17,
choi.vlsit16, choi.vlsit16}.
We characterize the impact of dwell time, i.e.,
idle time between consecutive program cycles, and environment temperature on
the retention loss speed and programming accuracy in 3D charge trap NAND flash
cells~\cite{luo.hpca18}.
The field (both academia and industry) is currently in much need of rigorous
experimental characterization and analysis of 3D NAND flash memory devices.

\textbf{Read Disturb Error Mitigation.} Prior work proposes to
mitigate read disturb errors by caching recently read data to avoid a read
operation~\cite{sugahara.patent14}. Prior work also proposes to mitigate read
disturb errors using an idea similar to remapping-based
refresh~\cite{cai.iccd12}, known as \emph{read reclaim}. The key idea of read
reclaim is to remap the data in a block to a new flash block, if the block has
experienced a high number of reads~\cite{ha.apsys13, ha.tcad16, kim.patent12,
frost.patent10}. To bound the number of read disturb errors, some flash vendors
specify a maximum number of tolerable reads for a flash block, at which point
read reclaim rewrites the data to a new block (just as is done for remapping-
based refresh).

Two mechanisms are currently being implemented within Yaffs (Yet Another Flash
File System) to handle read disturb errors, though they are not yet
available~\cite{yaffs.mitigation}.  The first mechanism is similar to read
reclaim~\cite{ha.apsys13}, where a block is rewritten after a fixed
number of page reads are performed to the block (e.g., 50,000 reads for an MLC
chip).  The second mechanism periodically inserts an additional read (e.g., a read every
256 block reads) to a page within the block, to check whether that page has
experienced a read disturb error, in which case the page is copied to a new
block.

Recent work proposes to remap read-hot pages to blocks configured as SLC, which
are resistant to read disturb~\cite{zhu2017alarm, liu2015read}. Ha et al.\
combine this read-hot page mapping technique with our $V_{pass}$ Tuning
technique and read reclaim~\cite{ha.tcad16} to further reduce read
disturb errors. This shows that the techniques proposed by prior work are
orthogonal to our read disturb mitigation techniques, and can be combined with
our work for even greater protection.

\textbf{Voltage Optimization.}
While the pass-through voltage optimization is specific to read disturb error
mitigation, a few works that propose optimizing the read reference
voltage have the same spirit~\cite{papa.glsvlsi2014, cai.iccd13,
cai.sigmetrics14}. Cai et al.\ propose a technique to calculate the optimal
read reference voltage from the mean and variance of the
threshold voltage distributions~\cite{cai.sigmetrics14}, which are characterized by the
read-retry technique~\cite{cai.date13}. The cost of such a technique is relatively
high, as it requires periodically reading flash memory
with all possible read reference voltages to discover the threshold
voltage distributions. Papandreou et al.\ propose to apply
a per-block close-to-optimal read reference voltage by periodically
sampling and averaging 6 OPTs within each block,
learned by exhaustively trying all possible read reference voltages~\cite{papa.glsvlsi2014}.
In contrast, ROR can find the actual optimal read reference
voltage at a much lower latency, thanks to the new findings and observations in
our DSN 2015 paper~\cite{cai.hpca15}. We already showed in our DSN 2015 paper that ROR greatly
outperforms naive read-retry, which is
significantly simpler than the mechanism proposed in~\cite{papa.glsvlsi2014}.

Recently, Luo et al.\ propose to accurately predict the optimal read
reference
voltage using an online flash channel model for each chip learned
online~\cite{luo.jsac16}. Cai et al.\ proposes a new technique called
$V_{pass}$ tuning, which tunes
the \emph{pass-through voltage}, i.e., a high reference voltage applied to
turn on unread cells in a block, to mitigate read disturb
errors~\cite{cai.dsn15}.
Du et al.\ proposes to tune the optimal read reference voltages for ECC
code soft decoding to improve ECC correction capability~\cite{du2017reducing}.
Fukami et al.\ proposes to use read-retry to improve the
reliability of chip-off forensic analysis of NAND flash memory
devices~\cite{fukami2017improving}.

\textbf{Error Recovery.} To our knowledge, no prior work other than
our DSN 2015 paper can recover the data from an uncorrectable error that is beyond
the error correction capability of ECC caused by read disturb~\cite{cai.dsn15}.
We have proposed a mechanism called RFR to opportunistically recover from
uncorrectable data retention errors~\cite{cai.dsn15, cai.procieee17, cai.procieee.arxiv17, cai.arxiv17}. RFR,
similar to RDR proposed in this work, identifies \emph{fast- and slow-leaking
cells}, rather than disturb-prone and disturb-resistant cells, and
probabilistically correct uncorrectable retention errors offline.

\chII{

\subsection{Read Disturb Errors in DRAM}
\label{sec:related:dramreaddisturb}

Commodity DRAM
chips that are sold and used in the field today exhibit read
disturb errors~\cite{kim.isca14}, also called \emph{RowHammer}-induced errors~\cite{mutlu.date17}, 
which are \emph{conceptually} similar to the read disturb
errors found in NAND flash memory.
Repeatedly accessing the same row in DRAM can cause
bit flips in data stored in adjacent DRAM rows. In order to
access data within DRAM, the row of cells corresponding
to the requested address must be \emph{activated} (i.e., opened for
read and write operations). This row must be \emph{precharged}
(i.e., closed) when another row in the same DRAM bank
needs to be activated. Through experimental studies on a
large number of real DRAM chips, we show that when a
DRAM row is activated and precharged repeatedly (i.e.,
\emph{hammered}) enough times within a DRAM refresh interval,
one or more bits in physically-adjacent DRAM rows can be
flipped to the wrong value~\cite{kim.isca14}.

We tested 129~DRAM modules manufactured by three major manufacturers
(A, B, and C) between 2008 and 2014, using an FPGA-based experimental DRAM
testing infrastructure~\cite{hassan.hpca17} (more detail on our experimental
setup, along with a list of all modules and their characteristics, can be found 
in our original RowHammer paper~\cite{kim.isca14}).  
Figure~\ref{fig:rowhammer-date} shows the
rate of RowHammer errors that we found, with the 129~modules that we tested
categorized based on their manufacturing date.
We find that 110 of our tested modules exhibit RowHammer errors, with the
earliest such module dating back to 2010.  In particular, we find that
\emph{all} of the modules manufactured in 2012--2013 that we tested are
vulnerable to RowHammer. Like with many NAND flash memory error 
mechanisms, especially read disturb, RowHammer
is a recent phenomenon that especially affects DRAM chips manufactured with more advanced
manufacturing process technology generations.

\begin{figure}[h]
  \centering
  \includegraphics[width=0.8\columnwidth]{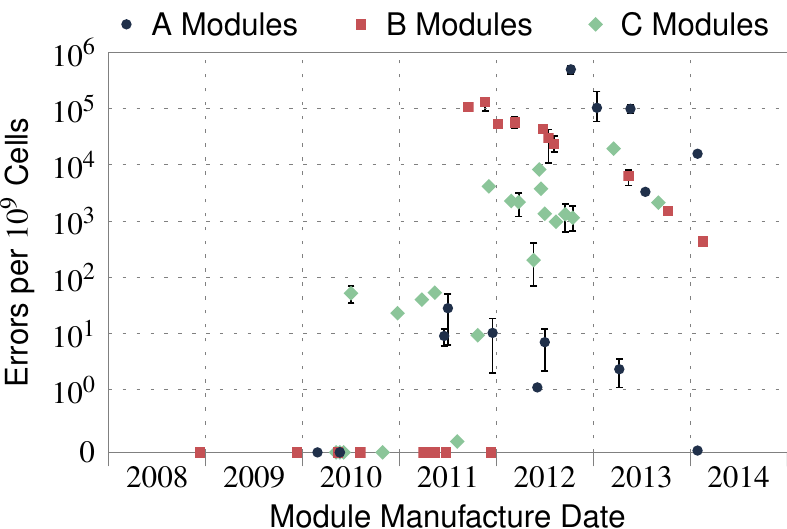}%
  \caption{RowHammer error rate vs.\ manufacturing dates of 129~DRAM
  modules we tested.  Reproduced from \cite{kim.isca14}.}%
  \label{fig:rowhammer-date}%
\end{figure}

Figure~\ref{fig:rowhammer-interference} shows the distribution of
the number of rows (plotted in log scale on the y-axis) within a DRAM module that flip the
number of bits along the x-axis, as measured for example DRAM 
modules from three different DRAM manufacturers~\cite{kim.isca14}.
We make two observations from the figure.  First, the number of bits
flipped when we hammer a row (known as the \emph{aggressor row}) can vary
significantly within a module.  Second, each module has a different 
distribution of the number of rows.
Despite these differences, we find that this DRAM failure mode
affects more than 80\% of the DRAM chips we tested~\cite{kim.isca14}.
As indicated above, this read disturb error mechanism in
DRAM is popularly called RowHammer~\cite{mutlu.date17}.

\begin{figure}[h]
  \centering
  \includegraphics[width=0.8\columnwidth]{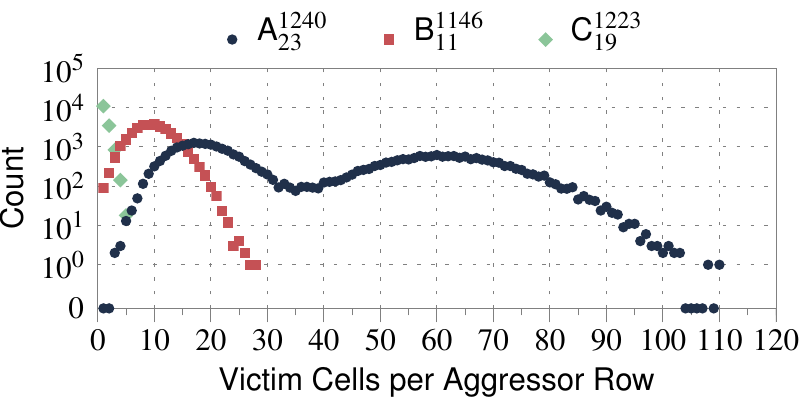}%
  \caption{Number of victim cells (i.e., number of bit errors) when an
  aggressor row is repeatedly activated, for three representative DRAM modules
  from three major manufacturers.  We label the modules in the format $X^{yyww}_n$, 
  where $X$ is the manufacturer (A, B, or C), $yyww$ is the manufacture year ($yy$) and
  week of the year ($ww$), and $n$ is the number of the selected module. 
  Reproduced from \cite{kim.isca14}.}%
  \label{fig:rowhammer-interference}%
\end{figure}


Various recent works show that RowHammer can be
maliciously exploited by user-level software programs to
(1)~induce errors in existing DRAM modules~\cite{kim.isca14, mutlu.date17}
and (2)~launch attacks to compromise the security of various
systems\chIII{~\cite{seaborn.blog15, mutlu.date17, seaborn.blackhat15, gruss.dimva16,
bosman.sp16, razavi.usenixsecurity16, vanderveen.ccs16, burleson.dac16,
xiao.usenixsecurity16, gruss.arxiv17, jang.systex17, poddebniak.crypto17}}.
For example, by exploiting the RowHammer read disturb
mechanism, a user-level program can gain kernel-level
privileges on real laptop systems~\cite{seaborn.blog15, seaborn.blackhat15}, take over a
server vulnerable to RowHammer~\cite{gruss.dimva16}, take over a victim
virtual machine running on the same system~\cite{bosman.sp16}, and
take over a mobile device~\cite{vanderveen.ccs16}. Thus, the RowHammer
read disturb mechanism is a prime (and perhaps the
first) example of how a circuit-level failure mechanism in
DRAM can cause a practical and widespread system security
vulnerability.

Note that various solutions to RowHammer exist~\cite{kim.isca14, mutlu.date17, kim.thesis15},
but we do not discuss them in detail here.  Our recent work~\cite{mutlu.date17} 
provides a comprehensive overview.  A very promising proposal is to modify
either the memory controller or the DRAM chip such that it probabilistically
refreshes the physically-adjacent rows of a recently-activated row, with very
low probability.  This solution is called \emph{Probabilistic Adjacent Row
Activation} (PARA)~\cite{kim.isca14}.  Our prior work shows that this low-cost, low-complexity
solution, which does not require any storage overhead, greatly closes the
RowHammer vulnerability~\cite{kim.isca14}.

The RowHammer effect in DRAM worsens as the manufacturing
process scales down to smaller node sizes~\cite{kim.isca14, mutlu.date17, mutlu.superfi14,mutlu.imw13}. 
More findings on RowHammer, along with extensive
experimental data from real DRAM devices, can be found in
our prior works~\cite{kim.isca14, mutlu.date17, kim.thesis15}.

\subsection{Errors in Emerging Memory Technologies}
\label{sec:related:emerging}

Emerging
nonvolatile memories\chII{~\cite{meza.weed13}}, such as \emph{phase-change memory} (PCM)~\cite{lee.isca09, qureshi.isca09, wong.procieee10, lee.ieeemicro10, zhou.isca09, lee.cacm10, yoon.taco14}, 
\emph{spin-transfer torque
magnetic RAM} (STT-RAM or STT-MRAM)~\cite{naeimi.itj13, kultursay.ispass13}, \emph{metal-oxide
resistive RAM} (RRAM)~\cite{wong.procieee12}, and \emph{memristors}~\cite{chua.tct71, strukov.nature08},
are expected to bridge the gap between DRAM and NAND-flash-memory-based SSDs, providing
DRAM-like access latency and energy, and at the same
time SSD-like large capacity and nonvolatility (and hence SSD-like
data persistence). 
While their underlying designs are different from DRAM and NAND flash memory,
these emerging memory technologies have been shown to exhibit similar types of
errors.

PCM-based devices are expected to have
a limited lifetime, as PCM can only endure a certain number
of writes~\cite{lee.isca09, qureshi.isca09, wong.procieee10}, similar to the P/E cycling errors in
SSDs (though PCM's write endurance
is higher than that of SSDs). PCM suffers from (1)~\emph{resistance
drift}~\cite{wong.procieee10, pirovano.ted04, ielmini.ted07}, where the resistance used to represent the value
becomes higher over time (and eventually can introduce a bit error),
similar to how charge leakage in NAND flash memory and
DRAM lead to retention errors over time; and
(2)~\emph{write disturb}~\cite{jiang.dsn14}, where the heat generated during the programming of one
PCM cell dissipates into neighboring cells and can change the value that is
stored within the neighboring cells, similar in concept
to cell-to-cell program interference in NAND flash memory.

STT-RAM suffers from (1)~\emph{retention failures}, where the value
stored for a single bit (as the magnetic orientation of the layer that 
stores the bit) can flip over time\chIII{~\cite{jog.dac12, sun.micro11, smullen.hpca11}};
and (2)~\emph{read disturb} (a
conceptually different phenomenon
from the read disturb in DRAM and flash memory), where
reading a bit in STT-RAM can inadvertently induce a write to
that same bit~\cite{naeimi.itj13}. 

Due to the nascent nature of emerging
nonvolatile memory technologies and the lack of availability of
large-capacity devices built with them, extensive and dependable
experimental studies have yet to be conducted on the reliability
of real PCM, STT-RAM, RRAM, and memristor chips.
However, we believe that error mechanisms conceptually or
abstractly similar to those for flash memory and DRAM are likely
to be prevalent in emerging technologies as well
(as supported by some recent studies~\cite{naeimi.itj13, jiang.dsn14, 
zhang.iccd12, khwa.isscc16, athmanathan.jetcas16, sills.vlsic15, sills.vlsit14}), 
albeit with different underlying mechanisms and error rates.

}


\section{Significance}


Our DSN 2015 paper~\cite{cai.dsn15} is the first openly-available work to
\chIII{(1)}~characterize the impact of read disturb errors on
commercially-available NAND flash memory devices,
\chIII{and (2)~propose novel solutions to the read disturb errors that minimize
them or recover them after error occurrence.}
We believe that our characterization
results, analyses, and mechanisms can have a wide impact on future research on
read disturb and NAND flash memory reliability.




\subsection{Long-Term Impact}

As flash devices continue to become more pervasive, there is renewed concern
\chIII{about the fewer} number of writes that these flash devices can
endure as they
continue to scale~\cite{cooke.fms07, yaffs.mitigation, ha.apsys13}.  This lower
\chIII{write} endurance is a result of the larger number of errors introduced
from \chIII{manufacturing process technology scaling, and the use of
multi-level cell technology}.
Today's planar NAND flash devices can \chIII{endure only on the order of} 100 program and
erase cycles~\cite{parnell.fms17} \chIII{without the assistance of aggressive error
mitigation techniques such as data refresh}~\cite{cai.iccd12,
luo.msst15}.

While there are
several solutions for other types of \chIII{NAND flash memory} errors, read disturb has in the past been
largely neglected because it has only become a significant problem at these
smaller process \chIII{technology nodes}~\cite{yaffs.mitigation, ha.apsys13}.
Our work
has the potential to change this \chIII{relative lack of attention to} read disturb
for several reasons:
\begin{itemize}
    \item We demonstrate on existing devices that read disturb is a \chIII{significant} problem
        today, and that it contributes a large number of errors \chIII{that further reduce NAND flash memory endurance}.
    \item We provide key \chIII{insights} as to \emph{why} these errors \chIII{occur},
        as well as why they will only worsen as \chIII{technology} scaling progresses.
    \item We show that it is possible to develop lightweight solutions that can
        alleviate the impact of read disturb.
\end{itemize}

Unfortunately, \chIII{unless error mitigation techniques for read disturb are
deployed in production NAND flash memory, read disturb} will
continue to negatively impact flash lifetime.  While \chIII{today's} 3D NAND flash
devices use larger process technologies that are less prone to read disturb
effects\chIII{~\cite{luo.thesis18, cai.arxiv17}},
future 3D NAND flash chips are expected to return to using smaller
process technologies \chIII{that remain} susceptible to read disturb, as manufacturers continue
to \chIII{aggressively increase} flash device densities~\cite{goda.iedm2012,
yoon.fms15, toshiba15nm}.
With flash devices expected to remain a large component of the storage market
for the foreseeable future, and with continued demand for higher flash
densities, we expect that our work on read disturb can inspire manufacturers
and researchers to adopt \chIII{effective} solutions to the read disturb problem.

The recovery mechanism that we propose, RDR, provides a new protective scheme
for data storage that people have \chIII{not} considered before.  \chIII
{Today, an increasingly larger volume of data is stored in data centers
belonging to cloud service providers, who must provide a strong guarantee of
data integrity for their end users.}
With flash storage continuing to expand in data
centers\chIII{~\cite{meza.sigmetrics15,schroeder2017reliability, narayanan.systor16}}, RDR (as
well as other recovery solutions that RDR might inspire) can reduce the 
probability of unrecoverable data loss for high-density storage.  In fact, the
availability of a recovery mechanism \chIII{like} RDR can also influence more data
centers to adopt flash memory for storage.

\subsection{New Research Directions}

In our \chIII{DSN 2015 paper~\cite{cai.dsn15}, we present} a number of \chIII{new} quantitative results on the
\chIII{impact of read disturb errors on NAND flash reliability}, as well as how several key factors
affect \chIII{the number of errors induced by read disturb}, such as the pass-through voltage, the number of
program/erase cycles, and \chIII{the} retention age.  
\chIII{Such a detailed characterization was not openly available in the past.}
We believe that by \chIII{releasing our characterization data}, 
researchers in both academia and industry will be able to \chIII{use the data to}
develop further mechanisms for read disturb recovery and mitigation.  In
addition, by exposing the importance of the read disturb problem in
contemporary \chIII{NAND} flash devices, we expect that our work will draw more attention
to the problem, and will inspire other researchers to further characterize
and understand the read disturb phenomenon.

\chIII{In fact, one of our recent
works builds on our DSN 2015 paper and shows that read disturb errors
can potentially cause security vulnerabilities in modern SSDs~\cite{cai.hpca17}.}

We also expect that RDR, our recovery approach, will inspire researchers to
design other data recovery mechanisms \chIII{for NAND flash memory} that also
leverage the intrinsic
properties of flash devices.  To our knowledge, our \chIII{new data} recovery
mechanism is the
first to do so, by discovering and exploiting the variation in read disturb
shifts that arise from the underlying process variation within a flash chip. 


\section{Conclusion}
\label{sec:conclusion}

We provide the first detailed experimental characterization of read
disturb errors for 2Y-nm MLC NAND flash memory chips. We find that bit errors
due to read disturb are much more likely to take place in cells with lower
threshold voltages, as well as in cells with greater wear.  We also find that
reducing the pass-through voltage can effectively mitigate read disturb errors.
Using these insights, we propose (1) a mitigation mechanism, called
\emph{$V_{pass}$ Tuning}, which dynamically adjusts the pass-through voltage for
each flash block online to minimize read disturb errors, and (2) an error
recovery mechanism, called \emph{Read Disturb Recovery}, which exploits the
differences in susceptibility of different cells to read disturb, to probabilistically
correct read disturb errors.
We hope that our characterization and analysis of the read disturb phenomenon
enables the development of other error mitigation and tolerance mechanisms,
which will become increasingly necessary as continued flash memory scaling leads
to greater susceptibility to read disturb. We also hope that our results will
motivate NAND flash manufacturers to add pass-through voltage controls to
next-generation chips, allowing flash controller designers to exploit our
findings and design controllers that tolerate read disturb more effectively.

\section*{Acknowledgments}

We thank the anonymous reviewers for their feedback. 
This work is partially supported by
the Intel Science and Technology Center, the CMU Data Storage
Systems Center, and NSF grants 0953246, 1065112, 1212962, and 1320531.

\bibliographystyle{IEEEtranS}
\bibliography{db}

\end{document}